# The orbit, mass, size, albedo, and density of (65489) Ceto/Phorcys: A tidally-evolved binary Centaur

W.M. Grundy[1], J.A. Stansberry[2], K.S. Noll[3], D.C. Stephens[4], D.E. Trilling[2], S.D. Kern[3], J.R. Spencer[5], D.P. Cruikshank[6], and H.F. Levison[5].

1. Lowell Observatory, 1400 W. Mars Hill Rd., Flagstaff AZ 86001.
2. Steward Observatory, University of Arizona, 933 N. Cherry Ave., Tucson AZ 85721.
3. Space Telescope Science Institute, 3700 San Martin Dr., Baltimore MD 21218.
4. Formerly at Dept. of Physics and Astronomy, Johns Hopkins University, Baltimore MD 21218; now at Dept. of Physics and Astronomy, Brigham Young University, N283 ESC Provo UT 84602.
5. Southwest Research Institute, 1050 Walnut St. #400, Boulder CO 80302.
6. NASA Ames Research Center, MS 245-6, Moffett Field CA 94035.

Primary contact:      Will Grundy
E-mail:               W.Grundy@lowell.edu
Voice:                928-233-3231
Fax:                  928-774-6296

Running head:        Binary Centaur (65489) Ceto/Phorcys
Manuscript pages:    25
Figures:             5
Tables:              5

**ABSTRACT**

Hubble Space Telescope observations of Uranus- and Neptune-crossing object (65489) Ceto/Phorcys (provisionally designated 2003 $FX_{128}$) reveal it to be a close binary system. The mutual orbit has a period of $9.554 \pm 0.011$ days and a semimajor axis of $1840 \pm 48$ km. These values enable computation of a system mass of $(5.41 \pm 0.42) \times 10^{18}$ kg. Spitzer Space Telescope observations of thermal emission at 24 and 70 μm are combined with visible photometry to constrain the system's effective radius ($109 \,^{+10}_{-11}$ km) and geometric albedo ($0.084 \,^{+0.021}_{-0.014}$). We estimate the average bulk density to be $1.37 \,^{+0.66}_{-0.32}$ g $cm^{-3}$, consistent with ice plus rocky and/or carbonaceous materials. This density contrasts with lower densities recently measured with the same technique for three other comparably-sized outer Solar System binaries (617) Patroclus, (26308) 1998 $SM_{165}$, and (47171) 1999 $TC_{36}$, and is closer to the density of the saturnian irregular satellite Phoebe. The mutual orbit of Ceto and Phorcys is nearly circular, with an eccentricity $\leqslant 0.015$. This observation is consistent with calculations suggesting that the system should tidally evolve on a timescale shorter than the age of the solar system.

## 1. Introduction

Objects following unstable, giant-planet crossing orbits in the outer solar system have been termed Centaurs[1] (e.g., Kowal et al. 1979; Elliot et al. 2005). They are thought to be bodies perturbed from the transneptunian region, evolving on timescales of the order of $\sim 10^7$ years to either being ejected from the solar system, suffering catastrophic impacts, roasting in the heat of the Sun as comets, or getting parked in cold-storage in the Oort cloud (e.g., Duncan and Levison 1997; Holman 1997; Levison and Duncan 1997; Horner et al. 2004). Centaurs offer a sample from the transneptunian region that is closer and more accessible to Earth-bound observers. They are less accessible than comets, but could have surfaces more pristine than those of comets which have spent more time closer to the Sun.

A typical member of the Centaur population has experienced numerous close encounters with one or more of the giant planets (e.g., Duncan et al. 1988; Levison and Duncan 1997). Such events might be expected to induce observable effects on the mutual orbits of binary Centaurs, when compared statistically with the mutual orbits of binary transneptunian objects (TNOs) in stable heliocentric orbits. For several years, Noll et al. (2004a, 2004b, 2006, 2007) have been carrying out a deep search for binaries among various dynamical classes of TNOs, most recently using the Advanced Camera for Surveys High Resolution Camera (ACS/HRC, Ford et al. 1996) aboard the Hubble Space Telescope (HST). An important result from this work is the finding that the dynamically cold "Classical" sub-population has relatively high binary rates compared with other transneptunian sub-populations (Stephens and Noll 2006). Recently, we (KSN, WMG, DCS, and HFL) have put greater emphasis on searching for binaries among the more dynamically excited sub-populations, including the Centaurs. Prior to its untimely demise, we used ACS/HRC to search for satellites around 22 Centaurs (as per the DES definition), finding companions for two of them. The first, (42355) Typhon/Echidna (provisionally designated 2002 $CR_{46}$), was discovered in 2006 January and a second, (65489) Ceto/Phorcys (provisionally designated 2003 $FX_{128}$), was discovered in 2006 April (Noll et al. 2006). The latter system is the subject of this paper.

---

[1] In this paper we use the Deep Ecliptic Survey (DES) definition of Centaurs as non-resonant objects with perihelia between the orbits of Jupiter and Neptune (Elliot et al. 2005). The Minor Planet Center defines transneptunian dynamical classes somewhat differently, and would classify some DES Centaurs, including (65489) Ceto/Phorcys, as members of the scattered disk (e.g., Gladman et al. 2007). The Committee on Small Body Nomenclature of the International Astronomical Union has recently adopted a naming convention for objects on unstable, non-resonant, giant-planet-crossing orbits with semimajor axes greater than Neptune's. Befitting their Centaur-like transitional orbits between TNOs and comets, they are to be named for other hybrid and shape-shifting mythical creatures. Some 30 objects are currently known to fall into this category, although only Ceto/Phorcys and (42355) Typhon/ Echidna have been named according to the new policy, so far. The names (announced without explanation in IAUC 8778) come from classical Greek mythology. The primary body is named for Ceto ($K\eta\tau o\varsigma$), an enormous sea monster, born of Gaia. English words like "cetacean" derive from her name. The secondary is named for Phorcys ($\Phi\acute{o}\rho\kappa\upsilon\varsigma$), her brother and husband, depicted as part man, part crab, and part fish. Together, they begot numerous terrible monsters, including the Gorgons.

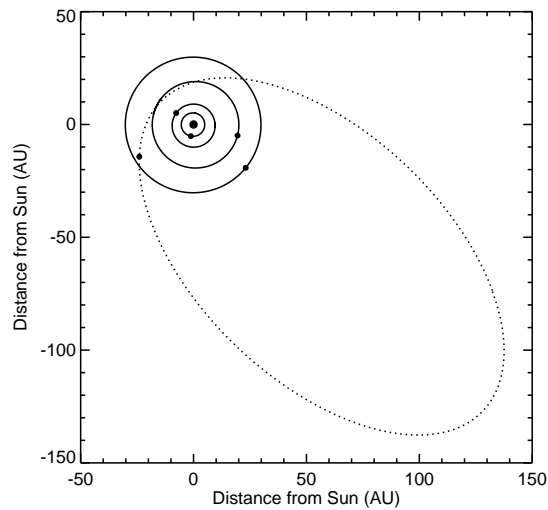

Fig. 1: Plan view of the Solar System, showing the heliocentric orbit of (65489) Ceto/Phorcys (dotted curve) compared with orbits of the four gas giants (solid curves), with points indicating positions as of 2007 July. The orbit crosses the orbits of both Uranus and Neptune, but considerable time is also spent well beyond 94 AU from the Sun, the distance at which Voyager 1 crossed the heliosphere termination shock (e.g., Decker et al. 2005)..

The initial discovery of (65489) Ceto/Phorcys was 2003 March 22 at the Palomar 1.2 m Schmidt camera by C. Trujillo and the Near-Earth Asteroid Tracking (NEAT) team (MPEC 2003-H33). It was soon identified in a series of pre-discovery images dating back to 1987, enabling its elongated heliocentric orbit (shown in Fig. 1) to be determined, with osculating heliocentric elements: semimajor axis $a_\odot = 102$ AU, inclination $i_\odot = 22°$, and eccentricity $e_\odot = 0.82$. The orbit crosses the orbits of both Uranus and Neptune and also extends far beyond the solar wind radiation environment to an aphelion of about 186 AU, well beyond the 94 AU distance where Voyager 1 crossed (or was crossed by) the heliosphere termination shock (e.g., Decker et al. 2005) with important implications for radiolytic processing of its surface (Cooper et al. 2003).

## 2. Hubble Space Telescope Observations and Analysis

The binary nature of (65489) Ceto/Phorcys was discovered in HST ACS/HRC images obtained 2006 April 11 UT (see Fig. 2), as part of Cycle 14 program 10514 (led by KSN). Details of the clear filter combination and dithering techniques used in program 10514 are described by Noll et al. (2006).

Follow up ACS/HRC images were obtained on May 6, 10, 30, and 31 UT as part of Cycle 14 program 10508 (led by WMG). The scheduling of these observations benefited from a statistical ranging Monte Carlo type of analysis inspired by the work of J. Virtanen and colleagues (e.g., Virtanen et al. 2001, 2003). This approach enabled us to optimize the timing of subsequent observations for satellite orbit determination, within the limitations of HST scheduling and the restricted visibility windows imposed by the two-gyro mode in which HST is currently operating. Our implementation of the technique is described in a forthcoming paper (Grundy et al. in preparation). The follow up observations used the *F606W* and *F814W* filters of the ACS/HRC camera. For each visit, eight 223 second exposures (four in each filter) were obtained, dithered over a four-point box pattern so that images with defects such as hot pixels or cosmic rays contaminating the flux of the binary system could be easily identified and excluded from further analysis.

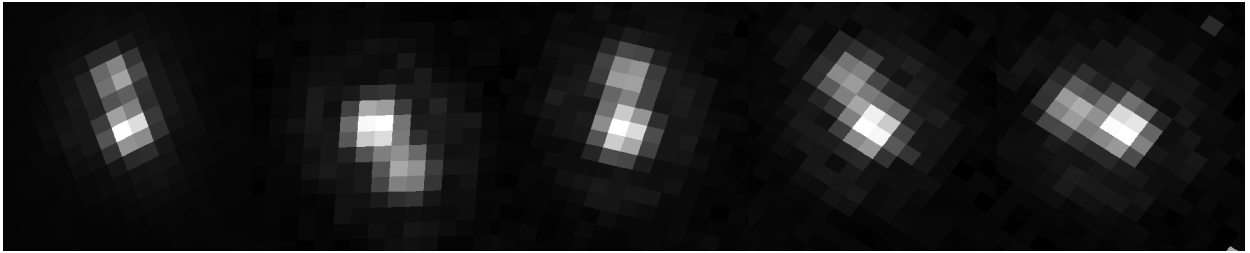

Fig. 2: Example Hubble Space Telescope ACS/HRC single frames of Ceto and Phorcys (centered on Ceto), taken 2006 April 11, May 5, May 10, May 30, and May 31 UT, from left to right. The HRC pixels are projected to sky-plane geometry with North up and East to the left, revealing the geometric distortion at the HRC focal plane. For scale, the mosaic of images is $2.0 \times 0.4$ arcsec. The left-most image was one of the binary discovery sequence of images, obtained through the *CLEAR* filter as part of program 10514. The other four images were obtained through the *F814W* filter as part of program 10508.

## 2.1. Photometry and astrometry

Data were processed through the standard CAL ACS pipeline at Space Telescope Science Institute (STScI), producing dark-subtracted, flat-fielded images (for details, refer to http://www.stsci.edu/hst/acs/documents/handbooks/cycle16). Astrometry and photometry were calculated from these images by fitting model point spread functions (PSFs) generated by Tiny Tim (Krist and Hook 2004) and correcting the results for the geometric distortion arising from the separation between the ACS camera and the optical axis of the telescope. Details of the binary-PSF fitting procedure are described in Stephens and Noll (2006). The *PyRAF* program *xytosky* in the *STSDAS* package (http://www.stsci.edu/resources/software_hardware) was used to translate the pixel coordinates found for Ceto and Phorcys to their true right ascension and declination values. Accurate separations and position angles for the binary pair could then be calculated, and measurements from the multiple images combined to get the average relative sky plane astrometry for each visit in Table 1.

Table 1. HST ACS/HRC observational circumstances and 1-$\sigma$ relative astrometry

| Average UT date and hour | $r$[a] | $\Delta$[a] | $g$[a] | $\Delta\alpha$[b] | $\Delta\delta$[b] |
|---|---|---|---|---|---|
| | | (AU) | (degrees) | | (arcsec) |
| 2006/04/11 21.691 | 27.753 | 26.785 | 0.542 | +0.0172 ± 0.002 | +0.0833 ± 0.003 |
| 2006/05/06 7.447 | 27.813 | 26.881 | 0.812 | −0.0475 ± 0.002 | −0.0687 ± 0.001 |
| 2006/05/10 4.510 | 27.822 | 26.912 | 0.916 | −0.0032 ± 0.001 | +0.0860 ± 0.002 |
| 2006/05/30 0.217 | 27.870 | 27.129 | 1.440 | +0.0362 ± 0.001 | +0.0747 ± 0.001 |
| 2006/05/31 1.692 | 27.873 | 27.143 | 1.466 | +0.0829 ± 0.002 | +0.0295 ± 0.002 |

The PSF model binary solutions were also used to calculate individual magnitudes for Ceto and Phorcys. To get the magnitude of one object, we subtracted the model PSF of the other from each image and then used the *PyRAF multidrizzle* task in the *STSDAS* package to remove the geometric distortion. For each resulting undistorted image, the total flux within a 3 pixel radius was measured and corrected to an aperture of 20 pixels (0.5 arcsec) computed for an undistorted model PSF. We then applied the Sirianni et al. (2005) AC05 (0.5 arcsec to ∞) aperture correction coefficients, the photometric zero point for each filter, and the charge transfer efficiency correction to get the final HST magnitude in each image.

To compare the HST magnitudes with ground based results and to help constrain the size and albedo of (65489) Ceto/Phorcys, transformation coefficients were derived to convert the *F606W* and *F814W* magnitudes to Johnson-Cousins *V* and *I* magnitudes. The ACS and Johnson-Cousins filters are not exact counterparts, so assumptions had to be made about the spectral energy distribution of Ceto/Phorcys. We applied a reddening coefficient to a solar spectrum until it had the same *F606W−F814W* color as (65489) Ceto/Phorcys. We then convolved this spectrum with the HST and Johnson-Cousins filters using the *synphot* task under *STSDAS* to get transformation coefficients specific to Ceto/Phorcys. These coefficients were used to transform the HST magnitudes of Ceto and Phorcys in each image to the Johnson-Cousins system. For each visit the *F606W*, *F814W*, *V*, and *I* magnitudes were averaged to produce Table 2.

Table 2. Photometry from Hubble Space Telescope ACS/HRC observations

| Observation | Ceto | | | | Phorcys | | | |
|---|---|---|---|---|---|---|---|---|
| Date (UT) | *F606W* | *F814W* | *V* | *I* | *F606W* | *F814W* | *V* | *I* |
| 2006/05/06 | 21.43 | 20.54 | 21.68 | 20.56 | 21.94 | 21.12 | 22.15 | 21.14 |
| 2006/05/10 | 21.35 | 20.53 | 21.57 | 20.55 | 21.94 | 21.10 | 22.18 | 21.13 |
| 2006/05/30 | 21.49 | 20.61 | 21.74 | 20.63 | 22.10 | 21.24 | 22.35 | 21.26 |
| 2006/05/31 | 21.49 | 20.63 | 21.72 | 20.65 | 22.07 | 21.24 | 22.29 | 21.26 |

Table note: Photometric 1-$\sigma$ uncertainties in *F606W* and *F814W* filters are about ±0.04 mag. Color transformation uncertainties inflate the *V* and *I* uncertainties to about ±0.06 mag.

Two filters were used in the follow-up observations to search for color differences between Ceto and Phorcys. The average *F606W−F814W* colors were 0.867 ± 0.028 and 0.836 ± 0.028 mag for Ceto and Phorcys, respectively. Ceto's measured color is slightly redder than that of Phorcys, but the difference is statistically insignificant. Our average *V−I* color for the system is 1.07 ± 0.04, corresponding to a spectral slope of 15.3 ± 1.5 (% rise per 100 nm relative to *V*), putting Ceto/Phorcys among the gray color clump of Centaurs, albeit at the red edge of that clump (e.g., Boehnhardt et al. 2003; Peixinho et al. 2003). Tegler et al. (2003) reported a *B−V* color for (65489) Ceto/Phorcys of 0.86 ± 0.03 and a *V−R* color of 0.56 ± 0.03, consistent with this result.

No lightcurve variability has been reported for this system and our data show no evidence for it either. The average magnitudes from the 4 visits reported in Table 2 differ by more than the 1-$\sigma$ uncertainties, but that variability can be attributed to geometry. After removing photometric effects of the changing distances from the Sun to the target $r$ and from the observer to the target $\Delta$ and also the effect of the phase angle $g$ (by assuming generic dark-object phase behavior as represented by the $H$ and $G$ system of Bowell et al. 1989, with $G = 0.15$), we find that the largest deviation from the 4-visit mean for either object or filter is only 1.5-$\sigma$ (for Ceto during the May 10 UT visit in *F606W*). With only four random epochs and $\pm 0.04$ mag photometric uncertainties, the *F606W* data can only rule out a 0.15 mag sinusoidal lightcurve at about the 1-$\sigma$ confidence level. However, the observed *F606W* outlier is not mirrored in the *F814W* filter, as would be expected if a lightcurve from an elongated shape were responsible, so noise is the likelier explanation for that point. At least one deviation as large as 1.5-$\sigma$ is to be expected among four sets of four measurements.

## 2.2. Orbit fitting and system mass

Our methods for fitting mutual orbits to relative astrometry of binaries have been described previously (Noll et al. 2004a, 2004b, 2006, 2007). We use the downhill simplex "amoeba" algorithm (Nelder and Mead 1965; Press et al. 1992) to iteratively adjust a set of seven orbital elements to minimize the $\chi^2$ statistic of residuals between observed and predicted sky plane relative positions of the primary and secondary, accounting for light time delays and relative motion between the observer and the binary pair. For previous orbit fits we had used the date and argument of periapsis as two of the seven fitted orbital elements. This choice had originally been made because the orbit of the satellite of 1998 WW$_{31}$ (Veillet et al. 2002) suggested to us that very high eccentricities might be the norm. Unfortunately, for circular orbits those two parameters become degenerate, and as eccentricity approaches zero, their use produces numerical instabilities which plagued our fits to the near-circular orbits of Pluto's satellites (Buie et al. 2006). Accordingly, we have since modified our orbit-fitting code to fit the following seven elements: period $P$, semimajor axis $a$, eccentricity $e$, inclination $i$, mean longitude at epoch $\epsilon$, longitude of ascending node $\Omega$, and longitude of periapsis $\varpi$. To estimate the uncertainty in each of these fitted elements, we used two different methods. First, we systematically varied each parameter around its best fit value, allowing the other six to adjust themselves to re-minimize $\chi^2$, creating slices through the seven dimensional $\chi^2$ space which enabling us to map out the 1-$\sigma$ confidence contour. Second, we fitted orbits to random sets of input astrometry data generated from the observed astrometry by adding Gaussian noise consistent with the astrometric uncertainties. The bundle of Monte Carlo orbits generated in this way provides an alternate measure of how well we have determined the orbital elements. Both methods gave very similar 1-$\sigma$ uncertainties on the fitted elements for the orbit of Phorcys around Ceto. The elements are tabulated in Table 3 and the projection of the orbit on the sky plane at the average time of the HST observations is shown in Fig. 3.

Table 3. Orbital elements and derived parameters with 1-$\sigma$ uncertainties or limits

| Parameter | | Orbit 1 ($\chi^2 = 2.3$) | Orbit 2 ($\chi^2 = 2.4$) |
|---|---|---|---|
| Fitted orbital elements | | | |
| Period (days) | $P$ | $9.557 \pm 0.008$ | $9.551 \pm 0.007$ |
| Semimajor axis (km) | $a$ | $1840 \pm 41$ | $1841 \pm 47$ |

| Parameter | | Orbit 1 ($\chi^2 = 2.3$) | Orbit 2 ($\chi^2 = 2.4$) |
|---|---|---|---|
| Eccentricity | $e$ | $\leq 0.013$ | $\leq 0.015$ |
| Inclination[a] (deg) | $i$ | $68.8 \pm 2.9$ | $116.6 \pm 3.0$ |
| Mean longitude[a] at epoch[b] (deg) | $\epsilon$ | $23.0 \pm 2.7$ | $53.5 \pm 5.2$ |
| Longitude of asc. node[a] (deg) | $\Omega$ | $105.5 \pm 3.7$ | $134.6 \pm 3.4$ |
| Longitude of periapsis[a] (deg) | $\varpi$ | $40 \pm 360$[c] | $70 \pm 360$[c] |
| Derived parameters | | | |
| System mass ($10^{18}$ kg) | $M_{sys}$ | $5.40 \pm 0.36$ | $5.42 \pm 0.42$ |
| Orbit pole right ascension[a] (deg) | $\alpha$ | $15.5 \pm 5.2$ | $44.6 \pm 5.2$ |
| Orbit pole declination[a] (deg) | $\delta$ | $21.2 \pm 4.0$ | $-26.7 \pm 4.1$ |

[a] Referenced to J2000 equatorial frame.

[b] The epoch is Julian date 2453880 (2006 May 24 12:00 UT).

[c] The uncertainty of $\pm 360°$ indicates that this parameter is unconstrained.

It can be difficult to distinguish between an orbit and its mirror image through the plane of the sky (e.g., Descamps 2005). Parallaxes from the differential motion of the Earth and the object will eventually enable a patient observer to break the ambiguity between an orbit and its mirror, as the observer's sky plane gradually rotates with respect to the orbit. More face-on orbits, with their smaller excursions of the satellite in the direction radial to the observer, are relatively slow to reveal which solution is the real one. At the time of our observations, the Ceto/Phorcys orbit was tilted about 28° from face-on. We list both solutions, noting that $\chi^2$ is slightly better for the first one, but does not exclude the second one. For both orbits, the longitude of periapsis $\varpi$ is unconstrained, in that it can take any value from 0 to 360° without pushing $\chi^2$ over the 1-$\sigma$ threshold. However, there is a shallow $\chi^2$ minimum for $\varpi$ at about 40° for Orbit 1 and about 70° for Orbit 2. Neither orbit will be oriented edge-on to the inner solar system within the next century, so use of mutual events to obtain more detailed information about the system geometry (e.g., Noll et al. 2007) is a distant future prospect.

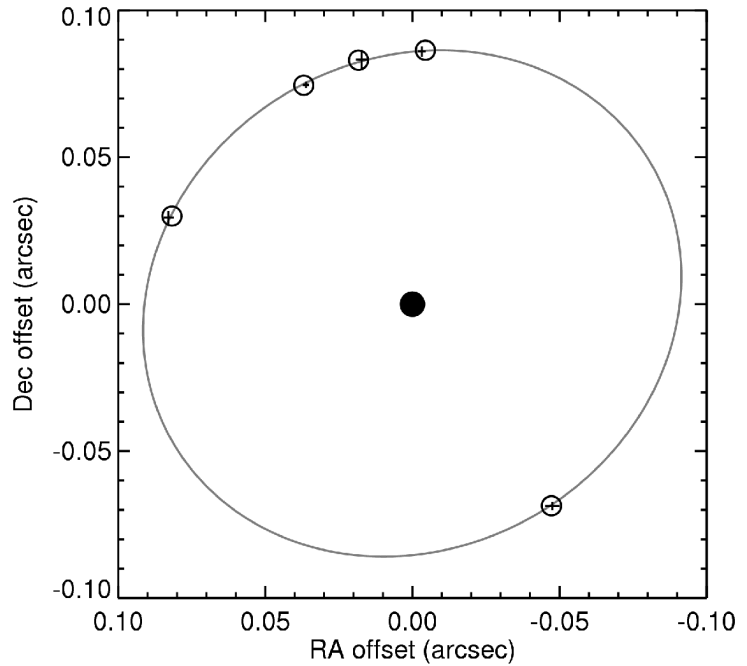

Fig. 3: Projection of the orbit of Phorcys (gray oval) relative to Ceto's location (black dot) onto the sky plane at the average time of the five HST observations. Points with 1-$\sigma$ error bars indicate relative astrometry between Ceto and Phorcys measured from the ensemble of dithered frames during each HST visit. Open circles are relative positions of Phorcys at the times of the observations as calculated from orbit 1 in Table 3. The spot at the origin and the open circles are scaled to the sizes of Ceto and Phorcys, respectively, derived in Section 4.

The mutual orbit of Ceto/Phorcys can be compared with orbits of satellites of other small outer Solar System bodies (see Noll et al. 2007 Table 2). It is noteworthy that this is the least massive system with a near-circular orbit by some three orders of magnitude. All known semimajor axes for transneptunian binaries are larger except for that of Typhon/Echidna (Grundy et al. in preparation) and all periods are longer except for that of Charon's orbit about Pluto. Although these characteristics sound exceptional, we do not believe this system to be unusual. Closer systems such as this one are simply more difficult to discovery and to study.

From the semimajor axis $a$ and period $P$, we can compute the mass of the combined system $M_{sys}$, according to

$$ M_{sys} = \frac{4\pi^2 a^3}{G P^2}, \tag{1} $$

where $G$ is the gravitational constant which we take to be $6.6742 \times 10^{-11}$ m$^3$ s$^{-2}$ kg$^{-1}$. Since we have two possible orbits, we adopt values of $P = 9.554 \pm 0.011$ days, $a = 1840 \pm 48$ km, and $M_{sys} = (5.41 \pm 0.42) \times 10^{18}$ kg to encompass the 1-$\sigma$ uncertainties from both solutions. The semimajor axis is measured to much lower fractional precision than the period is and it is raised to a higher power in the mass equation, so it is the dominant source of uncertainty in the Ceto/Phorcys system mass.

## 3. Spitzer Space Telescope Observations and Analysis

Spitzer Space Telescope observed (65489) Ceto/Phorcys in 2006 July with the Multiband Imaging Photometer for Spitzer (MIPS, Rieke et al. 2004) in its 24 and 70 μm bands, which have effective wavelengths of 23.68 and 71.42 μm. These observations were part of Cycle 3 program 30081 (led by JAS). Data were collected using the photometry observing template, which is tailored for photometry of point sources. The observatory tracked the target during the observations, although the motion was negligible relative to the size of the point-spread function (PSF) at these wavelengths. The target was imaged in both bands during two visits separated by 37 hours, over which time (65489) Ceto/Phorcys moved 38 arcsec relative to the background stars, allowing us to identify the target with certainty. Since the images from the two epochs have significant spatial overlap, we could use the image pairs to confirm that no strong background sources were present at the target position at either epoch and to subtract off the effect of weak background sources. The two visits would also help average over any possible thermal lightcurve variability. Images from the first epoch are shown in Fig. 4.

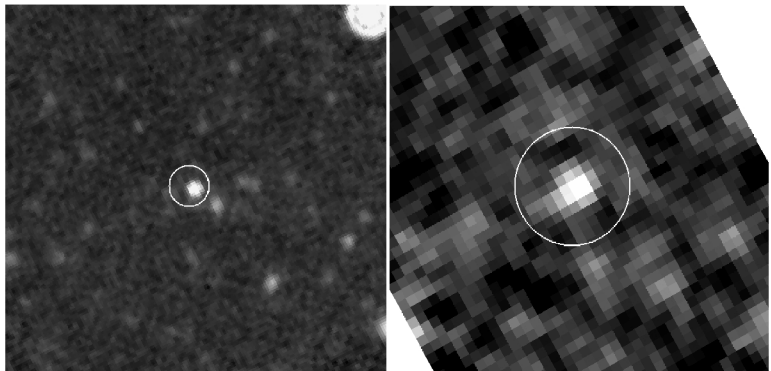

Fig. 4: Spitzer Space Telescope MIPS images of Ceto+Phorcys at 24 μm (left) and 70 μm (right) from 2006 July 18 UT. Each image is 180 arcsec square, and is oriented North up, East left. The circles are centered at the ephemeris position of the object, and have diameters equal to four times the FWHM of the Spitzer PSF in each band. While the object is not observed at exactly the ephemeris position, the offsets are consistent in the data from this and the other epoch, the object motion vector relative to the background objects is as predicted, and it is the only moving object visible at both epochs. Data from the second epoch were used to remove background sources from the first epoch, and vice versa.

We reduced the raw data and mosaicked them using the MIPS instrument team data analysis tools (Gordon et al. 2007). For the 24 μm data, basic processing included slope fitting, flat-fielding, and corrections for droop. All of these steps are currently implemented in the Spitzer Science Center (SSC) pipeline products. Additional corrections were made to remove readout offset (a jailbar pattern in the images), the effects of scattered light (which introduces a pointing-dependent background gradient and slightly degrades the sensitivity), and the application of a second-order flat field, derived from the data itself, to remove latents from previous observations.

The photometric repeatability of MIPS observations of moderately bright sources is better than 1% at 24 μm, and is 5% at 70 μm. The uncertainty in the absolute calibration of these bands is 4% and 8% respectively (Engelbracht et al. 2007; Gordon et al. 2007). For purposes of fitting models to our photometry, we use uncertainties that are the root-square-sum of the absolute calibration uncertainties and the measurement uncertainties determined from the images themselves. We adopt slightly larger 1-$\sigma$ calibration uncertainties, 5% and 10%, to account for uncertainties in color and aperture corrections. The widths of the filter bandpasses are about 25%, resulting in modest color corrections. We iteratively applied color corrections to our photometry, which converged to give a color temperature of 73.2 K, and color corrections of +2.6% and +8.6% at 24 and 70 μm, respectively (see Stansberry et al. 2007b). The uncertainties on the correction factors are perhaps a few percent of the factors themselves, and so are negligible for our study.

We measured the flux density of (65489) Ceto/Phorcys using 9.96 and 29.6 arcsec diameter apertures (about 4 and 3 native pixels) at 24 and 70 μm; the PSF full-width at half-max is 6.5 and 20 arcsec in those bands. The apertures were positioned at the center-of-light centroid. Mosaics were constructed using 1.245 arcsec and 4.925 arcsec pixels at 24 and 70 μm (about half the native pixel scale of those arrays). We applied aperture corrections of 1.91 and 1.85 to the 24 and 70 μm photometry to compute the total flux (see Gordon et al. 2007; Engelbracht et al. 2007 for details). Uncertainties in the aperture corrections are approximately 1%. Table 4 summarizes the circumstances of our Spitzer MIPS observations and the measured, color-corrected flux densities.

Table 4. Spitzer Space Telescope MIPS observational circumstances and thermal fluxes

| Average UT date and hour | $r$ | $\Delta$ | $g$ | 24 μm flux | 70 μm flux |
|---|---|---|---|---|---|
| | (AU) | | (degrees) | (mJy) | |
| 2006/07/18 10.23 | 27.990 | 27.662 | 1.99 | 1.47 ± 0.02[a] | 14.9 ± 1.0[a] |
| 2006/07/19 19.46 | 27.993 | 27.686 | 2.00 | 1.48 ± 0.02[a] | 14.5 ± 1.2[a] |
| Adopted | 27.991 | 27.674 | 2.00 | 1.473 ± 0.076[b] | 14.69 ± 1.84[b] |

[a] Errors reflect the signal to noise ratios in the observations, which were about 70 and 13, resulting from integrations of 900 and 2640 seconds, at 24 and 70 μm, respectively.

[b] The adopted 1-$\sigma$ flux errors include additional absolute calibration uncertainties.

## 4. Size, Albedo, and Density

The thermal fluxes measured by Spitzer can be used to constrain the size of Ceto/Phorcys and also the albedo if they can be combined with suitable visible photometry. Tegler et al. (2003) reported a $V$ magnitude of $20.70 \pm 0.03$ from Vatican Observatory 1.8 m telescope observations when the system was at heliocentric distance $r = 25.190$ AU, geocentric distance $\Delta = 24.298$ AU, and phase angle $g = 1.09°$. Assuming typical phase behavior for low albedo objects ($G = 0.15$ in the $H$ and $G$ system of Bowell et al. 1989; see also Sheppard and Jewitt 2002; Rousselot et al. 2005), we can convert the Tegler et al. $V$ photometry to absolute magnitude $H_V = 6.603 \pm 0.030$. Under the same assumptions, our HST photometry (in Tables 1 and 2) gives $H_V = 6.625 \pm 0.030$. Combining these, we adopt the weighted average $H_V = 6.614 \pm 0.021$.

Thermal fluxes are used to to constrain an object's size via a variety of thermal models (e.g., Lebofsky and Spencer 1989). We used the approach described by Stansberry et al. (2006), which takes advantage of the existence of visual photometry plus two well-separated thermal infrared wavelengths from Spitzer/MIPS to simultaneously fit the size, albedo, and beaming parameter $\eta$, using the standard thermal model (STM), which assumes instantaneous thermal equilibrium with sunlight illuminating a spherical surface. We also assumed unit emissivity. A wide range of real-world complications such as rotation, pole orientation, surface roughness, and thermal inertia can strongly influence observable thermal fluxes relative to what would be predicted by the simple STM, but in general, the effects of these complications look spectrally similar to the effect of changing the beaming parameter $\eta$. If $\eta$ is constrained by the data, the large model uncertainties in size and albedo which would otherwise flow from lack of knowledge of these other physical properties are greatly diminished. Using this model, we are able to match the observed thermal and visible photometry of (65489) Ceto/Phorcys with an effective radius $R_{\rm eff} = 109 \, {}^{+10}_{-11}$ km, along with beaming parameter $\eta = 0.77 \, {}^{+0.08}_{-0.09}$, and geometric albedo $A_{\rm p} = 0.084 \, {}^{+0.021}_{-0.014}$. Uncertainties on the derived parameters were determined by means of Monte Carlo methods. Thousands of random sets of input fluxes were generated, consistent with the observational uncertainties. Each set resulted in slightly different values of the derived parameters, creating a population of solutions from which 1-$\sigma$ uncertainties could be estimated. Errors from possible lightcurve variations are small enough to be neglected. For example, with thermal observations at two effectively random times, the 1-$\sigma$ upper limit 0.15 mag lightcurve mentioned earlier would contribute about a 1.4% average radius error, well below our ${}^{+9\%}_{-11\%}$ reported uncertainty.

The derived value of $\eta = 0.77 \, {}^{+0.08}_{-0.09}$ is on the low side of what has been seen for other TNOs and Centaurs, suggesting that Ceto and Phorcys have some combination of relatively low thermal inertias, slow rotation rates, especially rough surfaces, or their poles oriented toward the Sun. For comparison, Stansberry et al. (2007a) found $\eta$ values ranging from 0.6 to 2.3, with an average of 1.3, for a sample of 23 TNOs and Centaurs detected with good signal precision at both 24 and 70 μm wavelengths by Spitzer/MIPS.

Our $0.084 \, {}^{+0.021}_{-0.014}$ albedo of Ceto/Phorcys is consistent with albedos of other Centaurs. Stansberry et al. (2007a) report albedos for 11 other Centaurs (by the DES classification) detected with good signal precision at both 24 and 70 μm by Spitzer/MIPS. That sample has an average albedo of 0.06 and a standard deviation of 0.04. Considerable albedo diversity is also apparent among other small outer Solar System objects (e.g., Grundy et al. 2005; Cruikshank et al.

2007; Stansberry et al. 2007a), but correlations between albedo and dynamical class or color remain poorly established, except for a general trend of progressively lower average albedos from TNOs to Centaurs to comet nuclei, and a trend of higher albedos for the redder Centaurs (Stansberry et al. 2007a).

By assuming the albedos of Ceto and Phorcys are similar to one another, we can use the $0.584 \pm 0.030$ average magnitude difference between them to work out their individual radii, obtaining $R_{\mathrm{Ceto}} = 87\,^{+8}_{-9}$ km and $R_{\mathrm{Phorcys}} = 66\,^{+6}_{-7}$ km. Computing the total system volume from these two radii, and combining that with the total system mass from the mutual orbit, we obtain an average bulk density of $1.37\,^{+0.66}_{-0.32}$ g cm$^{-3}$. The assumption of equal albedos is consistent with the observation of indistinguishable colors for the two components, but we can also explore how sensitive we are to that assumption. Assume, for instance, that the albedo of Phorcys was actually twice that of Ceto. We would then have overestimated the average bulk density by a factor of 9%. If Ceto's albedo were double that of Phorcys, we would have underestimated the average bulk density by 2%. Unless the albedo contrast between Ceto and Phorcys is very extreme indeed, the density is relatively insensitive to our assumption of equal albedos. Our conclusions are comparably insensitive to possible lightcurve variations, as mentioned previously.

Bulk densities $\rho$ of icy objects are often interpreted in terms of a rock fraction

$$f_{\mathrm{rock}} \;=\; \frac{1 \,-\, \dfrac{\rho_{\mathrm{ice}}}{\rho}}{1 \,-\, \dfrac{\rho_{\mathrm{ice}}}{\rho_{\mathrm{rock}}}}\,, \tag{2}$$

where $\rho_{\mathrm{ice}}$ is taken to be 0.937 g cm$^{-3}$ for low temperature crystalline ice I$_{\mathrm{h}}$ (Lupo and Lewis 1979) and $\rho_{\mathrm{rock}}$ could range from 2.5 g cm$^{-3}$ for hydrated silicates to 3.5 g cm$^{-3}$ for anhydrous silicates (e.g., Schubert et al. 1986). For $\rho_{\mathrm{rock}} = 2.5$ g cm$^{-3}$, our average bulk density of Ceto/Phorcys yields a rock fraction $f_{\mathrm{rock}} = 0.51\,^{+0.29}_{-0.39}$, a value intermediate between the rock-rich compositions of Triton and Pluto and the more ice-rich compositions of some of the smaller satellites of Saturn. However, this rock fraction calculation only makes sense in the absence of void space, or of other possible interior components, such as carbonaceous materials.

We estimate the hydrostatic pressure at the cores of Ceto and Phorcys according to

$$P_0 \;=\; \frac{2}{3}\pi\,G\,\rho^2\,R^2 \tag{3}$$

(e.g., Stansberry et al. 2006) as $2.0\,^{+1.4}_{-0.6}$ and $1.1\,^{+0.9}_{-0.4}$ MPa, respectively. These pressures are insufficient to squeeze much void space out of cold, particulate H$_2$O ice. Laboratory experiments by Durham et al. (2005) show that at pressures in the 5 to 10 MPa range, appreciable compaction begins to occur in cold ice, but as much as 10% void space remains even at pressures as high as 100 MPa. If Ceto and Phorcys have remained cold since their accretion, we can anticipate relatively high fractions of void space within them, complicating interpretation of their densities in terms of ice and rock fractions.

Bulk densities are known for several much larger outer Solar System bodies, including Pluto and Charon ($2.03 \pm 0.06$ and $1.65 \pm 0.06$ g cm$^{-3}$, respectively; Buie et al. 2006), Eris ($2.26 \pm 0.25$ g cm$^{-3}$; Brown 2006), and (136108) 2003 EL$_{61}$ ($3.0 \pm 0.4$ g cm$^{-3}$; Rabinowitz et al.

2006). These higher densities are consistent with expectation for rock-rich and possibly differentiated objects which probably experienced warm interior temperatures at some time in their histories. Two other small binary TNOs have recently had their average bulk densities determined via the technique used in this paper. Stansberry et al. (2006) reported a bulk density of $0.50\,^{+0.30}_{-0.20}$ g cm$^{-3}$ for (47171) 1999 TC$_{36}$ and Spencer et al. (2006) reported a bulk density of $0.70\,^{+0.32}_{-0.21}$ g cm$^{-3}$ for (26308) 1998 SM$_{165}$. These much lower densities require significant void space, even for pure ice compositions. Low densities compared with likely bulk compositions have also been determined recently for the Trojan asteroid (617) Patroclus (Marchis et al. 2006) and for main belt asteroids (e.g., (90) Antiope; Descamps et al. 2007) as well as for much smaller comet nuclei (e.g., A'Hearn et al. 2005; Davidsson and Gutiérrez 2006). However, the irregular saturnian satellite Phoebe, with a size similar to that of Ceto+Phorcys, has a higher bulk density of $1.630 \pm 0.033$ g cm$^{-3}$ (Johnson and Lunine 2005). These diverse densities (shown in Figure 5) point to considerable variety in bulk compositions and interior structures among small outer Solar System bodies.

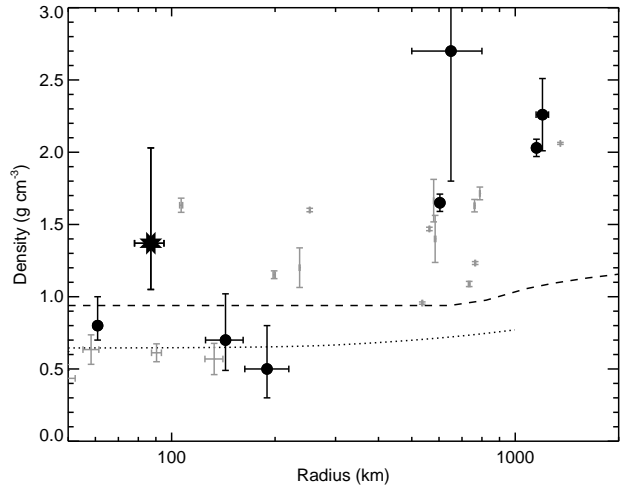

Fig. 5: Comparison of the average bulk density of (65489) Ceto/Phorcys (star) with densities for binary outer Solar System objects discussed in the text (black dots) and icy satellites of Saturn, Uranus, and Neptune (gray points, from Burns 1986 and McKinnon et al. 1995). For the binaries with unknown mass ratios, the radius shown is that of the primary, since it is likely to dominate the average density. The dashed curve is a theoretical bulk density for non-porous, pure $H_2O$ ice subject to self-compression at 77 K (Lupo and Lewis 1979). The dotted curve allows for porosity in pure, cold, granular $H_2O$ ice $I_h$, based on compaction data from Durham et al. (2005) as modeled by McKinnon et al. (2005). An expected trend of smaller objects having systematically lower densities than their larger counterparts (Lacerda and Jewitt 2007) is not cleanly supported by the present data for (65489) Ceto/Phorcys.

Apart from bulk compositional differences, thermal history is another potential source of density diversity among TNOs. Objects in the outer parts of the protoplanetary disk could initially accrete ice mostly in the amorphous phase, with a fluffy structure that includes considerable void space (e.g., Jenniskens et al. 1998). Subsequent thermal pulses, from whatever source (impacts, tidal dissipation, $Al^{26}$ decay, etc.), could crystallize amorphous ice. The crystallization would liberate additional heat as well as potentially reducing void space (e.g., Jewitt et al. 2007). The presence of varying amounts of $NH_3$ could also lead to considerable diversity of thermal evolution and thus density (e.g., Leliwa-Kopystyński and Kossacki 2000).

## 5. Tidal evolution

Since gravitational attraction falls off as the square of separation, Ceto and Phorcys both produce differential gravitational accelerations across the bodies of their partners, slightly distorting their equipotential figures. If their mutual orbit is eccentric, or either of their rotation states differs from the orbital rotation rate and orientation, this distortion will change over time, resulting in frictional dissipation of energy and leading to tidal evolution of their orbit and spin states. The rate of energy dissipation and tidal evolution depends on the specific tidal dissipation factor $Q$, which is a complicated function of interior structure and composition that describes the degree to which tidal friction is deposited in a given body. $Q$ is approximately equal to $1/(2\varepsilon)$, where $\varepsilon$ is the angle between the tidal bulge of the object and the line of centers between the two objects. Because Ceto and Phorcys have similar sizes and probably similar compositions, we assume that $Q_1 \approx Q_2 \approx Q$ (where subscripts 1 and 2 to refer to Ceto and Phorcys, respectively). Typical $Q$ values for solid planets and icy satellites range from 10 to 100 (e.g., Goldreich and Soter 1966; Yoder 1982; Hubbard 1984). We take $Q = 100$, to minimize dissipation and thus compute upper limits on tidal evolution timescales.

Small solid objects require a correction to $Q$ because their rigidity can be large compared to their self-gravity (Hubbard 1984). In this case, the small body may not be able to relax to hydrostatic equilibrium over a tidal cycle, but instead acts elastically. Thus, the deformation of the body's surface is smaller than it would be in hydrostatic equilibrium as elastic forces tend to maintain the body's shape. The correction to $Q$ takes the form

$$Q' = Q\left(\frac{1+19\mu}{2\,g\,\rho\,R}\right), \tag{4}$$

where $\mu$ is the rigidity, $\rho$ is the density, $R$ is the radius, and $g$ is the surface gravity (Goldreich and Soter 1966). Values adopted for use in our calculations are tabulated in Table 5. The rigidities of Ceto and Phorcys are, of course, unknown. If they are un-fractured monoliths, their rigidities could be as high as $10^{11}$ dyn cm$^{-2}$, intermediate between values for solid ice and solid rock (Gladman et al. 1996). In this extreme case, the correction to $Q$ would be large, with $Q' = 3 \times 10^6$. At the other extreme, if Ceto and Phorcys are unconsolidated "sand piles" their rigidities could be as low as $10^7$ dyn cm$^{-2}$ (Yoder 1982). In this case, the correction to $Q$ is small, giving $Q' = 300$. We use these bracketing values of $Q'$ to explore the range of possible timescales for tidal spin-locking and orbital evolution in the Ceto/Phorcys system.

Table 5. Physical parameters adopted.

| Parameter | | Ceto | Phorcys |
|---|---|---|---|
| Radius | $R$ | 87 km | 66 km |
| Mass | $M$ | $3.74 \times 10^{18}$ kg | $1.67 \times 10^{18}$ kg |
| Density | $\rho$ | 1.37 g cm$^{-3}$ | 1.37 g cm$^{-3}$ |
| Surface gravity | $g$ | 3.3 cm s$^{-2}$ | 2.5 cm s$^{-2}$ |

Table note: These values are derived from the effective radius $R_{\text{eff}}$, system mass $M_{\text{sys}}$, and magnitude difference, subject to the assumption of equal albedos and equal densities.

Over time, tidal effects can synchronize the spin rate of Phorcys to its orbital period (as for the Earth's Moon, which keeps one face oriented toward Earth) and, on a longer timescale, synchronize Ceto's spin rate as well (analogous to the Pluto-Charon system, where both members keep the same face toward each other). The timescale for spin-locking Ceto (slowing its spin rate to match the mutual orbital period) is given by

$$\tau_{\mathrm{spin},1} \;=\; \frac{Q'\,\Delta\,\omega_1 M_1 a^6}{G\,M_2^2 R_1^3}, \tag{5}$$

where subscripts 1 and 2 refers to Ceto and Phorcys, respectively, $M$ is mass, $R$ is radius, and $\Delta\omega$ is the difference between an initial angular rotation rate and the final, synchronized rate (Goldreich and Soter 1966; Trilling 2000). The timescale for spin-locking Phorcys can be found simply by swapping all the subscripts in Equation 5. A typical rotation period for TNOs is around 10 hours (e.g., Trilling and Bernstein 2006). The current orbital period is much longer than this, so we set $\Delta\omega_1$ (and $\Delta\omega_2$) to correspond to this rate (one could equally start with the breakup rotation rate, in the 2 to 3 hour period range). We find that the spin-locking timescale for Ceto ranges from $10^6$ to $10^{10}$ years and the timescale for spin-locking Phorcys is about a factor of three shorter. We conclude that unless these bodies are rigid monoliths, Phorcys should be spin-locked to Ceto and Ceto to Phorcys as well, or nearly so. (Recall that if $Q$ is smaller than our assumed maximum value of 100, the spin-lock timescale decreases, increasing the likelihood of spin-locking.)

The eccentricity of the Ceto/Phorcys mutual orbit can also evolve due to tides raised on either object by the other. Following Goldreich and Soter (1966) we can write the contributions of the two terms as

$$\frac{de}{dt} \;=\; \left(\frac{de}{dt}\right)_1 + \left(\frac{de}{dt}\right)_2, \tag{6}$$

where $(de/dt)_1$ accounts for the effect of tides raised on the primary by the secondary and $(de/dt)_2$ accounts for tides raised on the secondary by the primary. The two contributions can be expressed as

$$\frac{1}{e}\left(\frac{de}{dt}\right)_1 \;=\; \frac{171}{16}\sqrt{\frac{G}{M_1}}\left(\frac{R_1^5 M_2}{Q'\,a^{6.5}}\right)\mathrm{sign}\left(2\,\omega_1 - 3\,n_2\right) \tag{7}$$

where $n_2$ is the mean orbital motion of the secondary and

$$\frac{1}{e}\left(\frac{de}{dt}\right)_2 \;=\; -\frac{63}{4}\sqrt{G\,M_1^3}\left(\frac{R_2^5}{Q'\,M_2\,a^{6.5}}\right). \tag{8}$$

Employing our previous result that Ceto and Phorcys should be spin-locked (or nearly so), $\omega_1$ should be equal (or close) to $n_1$ so $(de/dt)_1$ is negative, implying that $de/dt$ must also be negative (since $(de/dt)_2$ is always negative)[2]. A negative $de/dt$ indicates circularization of the orbit.

Using the values for Ceto and Phorcys in Table 5, we find

$$\frac{1}{e}\frac{de}{dt} \;\approx\; -\frac{7\times10^{-12} + 1.4\times10^{-11}}{Q'}\mathrm{sec}^{-1} \tag{9}$$

Equating $e$ with $-de$ and solving for $dt$ to estimate the timescale to reduce an initial eccentricity to zero, we find that the circularization timescale ranges from $5\times10^5$ to $5\times10^9$ years for the previously discussed range of $Q'$.  Again, unless Ceto and Phorcys are rigid monoliths, this timescale is much less than the age of the Solar System (and, presumably, the age of the Ceto/Phorcys binary) so we expect the system to have lost most (or all) of its primordial orbital eccentricity.

---

2  For bodies of equal density, $(de/dt)_2/(de/dt)_1 \approx 1.5\ (R_1/R_2)$, so $(de/dt)_2$ always dominates, and the first term on the right hand side of Equation 6 is often ignored (e.g., Rasio et al. 1996; Trilling 2000). Thus, even when $P_1 < 2/3\ P_2$, giving positive $(de/dt)_1$, $|(de/dt)_1|$ can never be larger than $|(de/dt)_2|$, so $de/dt$ will always be negative.

The average density of (65489) Ceto/Phorcys is relatively high compared with two other small TNOs with eccentric mutual orbits (26308) 1998 $SM_{165}$ (Spencer et al. 2006) and (47171) 1999 $TC_{36}$ (Stansberry et al. 2006), leading us to ask whether the tidal evolution of the Ceto/Phorcys system could have driven internal temperatures high enough to enable ductile compaction and loss of void space.  What temperature would be required is not clear, but for crystalline ice $I_h$, it must exceed 120 K, since ice $I_h$ remains brittle and can maintain its pore space up to at least that temperature (e.g., Durham et al. 2005).  This temperature is considerably higher than equilibrium temperatures in the Kuiper belt.  Energy from both circularization and spin-locking would contribute to internal heating.  The specific energy (i.e., per unit mass) available from a change in orbital semimajor axis is

$$\frac{\Delta E_{circ}}{M} \ = \ \frac{G M_1}{2}(a_{initial}^{-1} - a_{final}^{-1}) \,, \tag{10}$$

where $a_{initial}$ and $a_{final}$ are the initial and final semimajor axes.  If the orbit changes by ~10% in the process of circularization, then the total energy deposited is $GM_1/(20a_f)$, and we may take $a_f$ as the presently observed value of $a$.  Rotational kinetic energy of a spinning body is $\frac{1}{2}I\omega^2$ where $\omega$ is the angular rotation rate and $I = \frac{2}{5}MR^2$ is the moment of inertia for a homogeneous spherical body.  The specific energy deposited in each body via spin locking is thus

$$\frac{\Delta E_{spin}}{M} \ = \ \frac{R^2}{5}(\omega_{initial}^2 - \omega_{final}^2) \,, \tag{11}$$

where $\omega_{initial}$ and $\omega_{final}$ are the initial and final rotation rates.  Taking the initial spin periods to be 10 hours and final spin periods to be 9.5 days, we obtain an energy input smaller than that from circularization by a factor of a few.  Combining the above parts, the total specific energy deposited in the system is

$$\frac{\Delta E}{M} \ = \ \frac{\Delta E_{circ}}{M} + \frac{\Delta E_{spin}}{M} \ = \ \frac{G M_1}{20\,a} + \frac{(R_1^2 + R_2^2)(\omega_{initial}^2 - \omega_{final}^2)}{5} \,, \tag{12}$$

which is approximately $10^5$ erg $g^{-1}$.  Lunine and Tittemore (1993) present an alternate derivation of the tidal heating of a synchronously rotating satellite.  Rewriting their expression for energy deposition, we have

$$\frac{dE}{dt} \ = \ \frac{63}{4}\sqrt{G M_1^3}\left(\frac{R_2^5}{M_2\,a^{6.5}\,Q'_2}\right)\left(\frac{G M_1 M_2 e^2}{a}\right) \tag{13}$$

which gives $dE/dt$ ranging from $10^{10}$ to $10^{14}$ erg $s^{-1}$ for our bracketing $Q'$ assumptions.  The total energy input is independent of rigidity and is around $10^{27}$ erg, or $10^5$ erg $g^{-1}$, in agreement with our derivation above.

The heat capacity of cold water ice $I_h$ depends on temperature, ranging from $3 \times 10^6$ erg $g^{-1}$ $K^{-1}$ at 30 K to over $10^7$ erg $g^{-1}$ $K^{-1}$ at 60 K (Giauque and Stout 1936), so even if all $10^5$ erg $g^{-1}$ of heat were deposited instantaneously instead of over a $10^6$ year or longer timescale, the temperature rise would be much less than a degree, far too small for viscous creep to assist compaction and loss of pore space. Comparable arguments can be made to show that heat tidally deposited by any close encounter with a giant planet which does not completely disrupt the system will also produce a negligible temperature rise. However, if tidal heating were highly localized, as might happen if energy dissipation was confined to the intersections of large scale internal fractures, the local temperature rise could be higher.

That the Ceto/Phorcys system is expected to be tidally spin-locked and circularized suggests that the spin axes of both objects could also have become aligned with the orbit pole. Such a configuration would help explain why no lightcurve has been reported for this object, and why there is no evidence for it in our ACS/HRC photometry in Table 2 nor in our Spitzer/MIPS photometry in Table 4. If the spin axes are aligned with the orbit pole, which was tilted about 28° from the line of sight from Earth at the time of the observations (as seen in Fig. 3), a lightcurve observable when the system was near equator-on would be reduced in amplitude by the current geometry by a factor of sin(28°), equal to 0.47. A more pole-on configuration is also consistent with the small beaming parameter $\eta$. Smaller values of $\eta$ correspond to warmer surface temperatures, as would be expected when observing a hemisphere mostly in continuous sunlight. Unfortunately, we cannot use the assumption of aligned spin orientations to constrain the thermal inertia of the surfaces of Ceto and Phorcys, since observable thermal flux becomes independent of thermal inertia in pole-on geometry (e.g., Lebofsky and Spencer 1989).

## 6. Effect of encounters

In its current heliocentric orbit, Ceto/Phorcys is most likely to encounter Uranus, since its perihelion is just inside the orbit of Uranus and its argument of perihelion is near 180° (see Fig. 1). Tidal disruption becomes probable for planetary encounters with closest approach distances within

$$r_{\mathrm{td}} \;=\; a \left( \frac{3\,M_{\mathrm{planet}}}{M_{\mathrm{sys}}} \right)^{1/3}, \tag{14}$$

where $a$ is the binary's semimajor axis and $M_{\mathrm{planet}}$ and $M_{\mathrm{sys}}$ are the masses of the planet and the binary system, respectively (Agnor and Hamilton 2006). For encounters between Ceto/Phorcys and Uranus, $r_{\mathrm{td}}$ is about 0.0045 AU, equivalent to 26 times the radius of Uranus or 360 times the Ceto/Phorcys separation $a$. For an encounter with Jupiter, $r_{\mathrm{td}}$ would be 0.013 AU, or 1000 times $a$. The probability that Ceto/Phorcys would have had an encounter closer than $r_{\mathrm{td}}$ with any of the giant planets on its way to being perturbed into its current orbit is not known, but by analogy with (42355) Typhon/Echidna (Noll et al. 2006), it is perhaps not surprising that the Ceto/Phorcys binary survived the sequence of planetary encounters which put it on its present-day orbit. Although these encounters are unlikely to have been close enough to disrupt the binary, as evidenced by the simple fact of its existence, they could possibly be expected to have had some effect on the eccentricity of the binary orbit. If so, a future survey of comparable TNBs with orbits sampling the range from zero to non-zero eccentricity may enable statistical limits to be placed on their $Q$ values.

Trilling (2007) explores the possibility of binary disruption by encounters with other

TNOs. For Ceto/Phorcys, the most likely time for such an encounter to have taken place would have been prior to perturbation into a short-lived Centaur orbit. The "hardness" $H$ of the binary can be expressed as a function of the mass $M_3$ of a third body being encountered as

$$H(M_3) = \frac{G M_1 M_2}{2 a M_3 \sigma_v^2} \tag{15}$$

(Binney and Tremaine 1987; Trilling 2007) where $M_1$ and $M_2$ are the masses of Ceto and Phorcys (assuming both have the same density), $a$ is the semimajor axis of their mutual orbit, and $\sigma_v$ is the velocity dispersion of potential interlopers, which is unknown, because we do not know what heliocentric orbit the system was on prior to its becoming a Centaur. For $H(M_3)$ greater than unity, the system is "hard" against disruption by interlopers of mass $M_3$. Assuming a velocity dispersion of 1 km s$^{-1}$ (Jewitt et al. 1996), we find that $H(M_3)$ is unity for interlopers in the 3 km size range, assuming an average density of 1 g cm$^{-3}$. For smaller interlopers, the Ceto/Phorcys system is "hard" but for larger interlopers, it is "soft" and could be disrupted. Durda and Stern (2000) estimated rates of impacts in the Kuiper belt as a function of impactor size, finding that on average, an object the size of Ceto+Phorcys should be struck by one interloper of this size over the age of the solar system. Additional close, but non-impacting encounters would also be expected. A possible consequence could be that binaries at least as widely separated as Ceto and Phorcys (i.e., all currently known TNBs other than the near-contact binaries) represent a remnant population which has lost many of its members over the age of Solar System, but that even tighter binaries could be immune to disruption in the present collisional environment, and so could reveal the primordial binary abundance. This argument supports the ideas of Goldreich et al. (2002) and Petit and Mousis (2004) that closer binaries could be considerably more abundant than the comparatively widely separated ones currently being discovered by ground-based and even HST searches (e.g., Kern and Elliot 2006; Noll et al. 2007).

## 7. Conclusion

Combined Hubble Space Telescope and Spitzer Space Telescope observations reveal the Centaur (65489) Ceto/Phorcys to be a binary system with a separation of $1840 \pm 48$ km. The average geometric albedo is $0.084^{+0.021}_{-0.014}$, consistent with radiometrically determined albedos of other Centaurs. If both components have equal albedos, as suggested by their shared colors, the radii of Ceto and Phorcys are $87^{+8}_{-9}$ and $66^{+6}_{-7}$ km, respectively. The average bulk density of the system is then $1.37^{+0.66}_{-0.32}$ g cm$^{-3}$, higher than recent densities for other comparably-sized binary outer Solar System objects, but smaller than the densities of much larger icy worlds such as Pluto, Triton, and Eris. Despite probable recent encounters with gas giants, the mutual orbit of Ceto and Phorcys is nearly circular, consistent with calculations showing that the system should become tidally circularized and synchronized on timescales shorter than the age of the Solar System.

## Acknowledgments

This work is based in part on NASA/ESA Hubble Space Telescope Cycle 14 program 10508 and 10514 observations. Support for these programs was provided by NASA through a grant from the Space Telescope Science Institute (STScI), which is operated by the Association of Universities for Research in Astronomy, Inc., under NASA contract NAS 5-26555. We are especially grateful to Tony Roman and Tricia Royle at STScI for their rapid reactions in scheduling the HST follow-up observations. STSDAS and PyRAF are products of STScI. This work is also based in part on Spitzer Space Telescope Cycle 3 program 30081 observations. Spitzer is operated by the Jet Propulsion Laboratory, California Institute of Technology under a contract with NASA through an award issued by JPL/Caltech. We thank two anonymous reviewers for numerous constructive suggestions on how to improve this manuscript. Finally, we are grateful to the free and open source software communities for empowering us with many of the tools used to complete this project, notably Linux, the GNU tools, OpenOffice.org, MySQL, FVWM, Python, IRAF, STSDAS, PyRAF, and TkRat.

## REFERENCES

Agnor, C.B., and D.P. Hamilton 2006. Neptune's capture of its moon Triton in a binary-planet gravitational encounter. *Nature* **441,** 192-194.

A'Hearn, M.F., M.J.S. Belton, W.A. Delamere, J. Kissel, K.P. Klaasen, L.A. McFadden, K.J. Meech, H.J. Melosh, P.H. Schultz, J.M. Sunshine, P.C. Thomas, J. Veverka, D.K. Yeomans, M.W. Baca, I. Busko, C.J. Crockett, S.M. Collins, M. Desnoyer, C.A. Eberhardy, C.M. Ernst, T.L. Farnham, L. Feaga, O. Groussin, D. Hampton, S.I. Ipatov, J.Y. Li, D. Lindler, C.M. Lisse, N. Mastrodemos, W.M. Owen Jr., J.E. Richardson, D.D. Wellnitz, and R.L. White 2005. Deep Impact: Excavating comet Tempel 1. *Science* **310,** 258-264.

Binney, J., and S. Tremaine 1987. *Galactic dynamics.* Princeton Univ. Press, Princeton NJ.

Boehnhardt, H., A. Barucci, A. Delsanti, C. de Bergh, A. Doressoundiram, J. Romon, E. Dotto, G. Tozzi, M. Lazzarin, S. Fornasier, N. Peixinho, O. Hainaut, J. Davies, P. Rousselot, L. Barrera, K. Birkle, K. Meech, J. Ortiz, T. Sekiguchi, J.I. Watanabe, N. Thomas, and R. West 2003. Results from the ESO large program on transneptunian objects and centaurs. *Earth, Moon, and Planets* **92,** 145-156.

Bowell, E., B. Hapke, D. Domingue, K. Lumme, J. Peltoniemi, and A. Harris 1989. Application of photometric models to asteroids. In *Asteroids II*, Eds. R.P. Binzel, T. Gehrels, and M.S. Matthews, Univ. of Arizona Press, Tucson AZ.

Brown, M.E. 2006. The largest Kuiper belt objects. Invited talk at AAS/DPS meeting, Pasadena CA, 2006 October 9-13.

Buie, M.W., W.M. Grundy, E.F. Young, L.Y. Young, and S.A. Stern 2006. Orbits and photometry of Pluto's satellites: Charon, S/2005 P1, and S/2005 P2. *Astron. J.* **132,** 290-298.

Burns, J.A. 1986. Some background about satellites. In *Satellites*, Eds. J.A. Burns, and M.S. Matthews, Univ. of Arizona Press, Tucson AZ.

Cooper, J.F., E.R. Christian, J.D. Richardson, and C. Wang 2003. Proton irradiation of Centaur, Kuiper belt, and Oort cloud objects at plasma to cosmic ray energy. *Earth, Moon, and Planets* **92,** 261-277.

Cruikshank, D.P., M.A. Barucci, J.P. Emery, Y.R. Fernández, W.M. Grundy, K.S. Noll, and J.A.

Stansberry 2007. Physical properties of trans-neptunian objects. In *Protostars and Planets V*. Eds. B. Reipurth, D. Jewitt, and K. Keil, Univ. of Arizona Press, Tucson AZ.

Davidsson, B.J.R., and P.J. Gutiérrez 2006. Non-gravitational force modeling of Comet 81P/Wild 2 I. A nucleus bulk density estimate. *Icarus* **180,** 224-242.

Descamps, P. 2005. Orbit of an astrometric binary system. *Celestial Mech. and Dynamical Astron.* **92,** 381-402.

Descamps, P., F. Marchis, T. Michalowski, F. Vachier, F. Colas, J. Berthier, M. Assafin, P.B. Dunckel, M. Polinska, W. Pych, D. Hestroffer, K.P.M. Miller, R. Vieira-Martins, M. Birlan, J.-P. Teng-Chuen-Yu, A. Peyrot, B. Payet, J. Dorseuil, Y. Léonie, and T. Dijoux 2007. Figure of the double asteroid 90 Antiope from adaptive optics and lightcurve observations. *Icarus* **187,** 482-499.

Decker, R.B., S.M. Krimigis, E.C. Roelof, M.E. Hill, T.P. Armstrong, G. Gloeckler, D.C. Hamilton, and L.J. Lanzerotti 2005. Voyager 1 in the foreshock, termination shock, and heliosheath. *Science* **309,** 2020-2024.

Duncan, M.J., and H.F. Levison 1997. A scattered disk of icy objects and the origin of Jupiter-family comets. *Science* **276,** 1670-1672.

Duncan, M., T. Quinn, and S. Tremaine 1988. The origin of short-period comets. *Astrophys. J.* **328,** L69-L73.

Durda, D.D., and S.A. Stern 2000. Collision rates in the present-day Kuiper belt and Centaur regions: Applications to surface activation and modification on Comets, Kuiper belt objects, Centaurs, and Pluto-Charon. *Icarus* **145,** 220-229.

Durham, W.B., W.B. McKinnon, and L.A. Stern 2005. Cold compaction of water ice. *Geophys. Res. Lett.* **32,** L18202.1-5.

Elliot, J.L., S.D. Kern, K.B. Clancy, A.A.S. Gulbis, R.L. Millis, M.W. Buie, L.H. Wasserman, E.I. Chiang, A.B. Jordan, D.E. Trilling, and K.J. Meech 2005. The Deep Ecliptic Survey: A search for Kuiper belt objects and Centaurs. II. Dynamical classification, the Kuiper belt plane, and the core population. *Astron. J.* **129,** 1117-1162.

Engelbracht, C.W., M. Blaylock, K.Y.L. Su, J. Rho, and G.H. Rieke 2007. Absolute calibration and characterization of the Multiband Imaging Photometer for Spitzer. I. The stellar calibrator sample and the 24 μm calibration. *Publ. Astron. Soc. Pacific* (submitted).

Ford, H.C., and the ACS Science Team 1996. The Advanced Camera for the Hubble Space Telescope. In *Space Telescopes and Instruments IV*. Eds. P. Bely and J. Breckinridge, SPIE **2807,** 184-196.

Giauque, W.F., and J.W. Stout 1936. The entropy of water and the third law of thermo-dynamics: The heat capacity of ice from 15 to 273°K. *J. Amer. Chem. Soc.* **58,** 1144-1150.

Gladman, B., D.D. Quinn, P. Nicholson, and R. Rand 1996. Synchronous locking of tidally evolving satellites. *Icarus* **122,** 166-192.

Gladman, B., B.G. Marsden, and C. VanLaerhoven 2007. Nomenclature in the outer Solar System. In *Kuiper Belt*. Eds. A. Barucci, H. Boehnhardt, D. Cruikshank, and A. Morbidelli.

Univ. of Arizona Press, Tucson AZ (in press).

Goldreich, P., and S. Soter 1966. Q in the solar system. *Icarus* **5,** 375-389.

Goldreich, P., Y. Lithwick, and R. Sari 2002. Formation of Kuiper-belt binaries by dynamical friction and three-body encounters. *Nature,* **420,** 643-646.

Gordon, K.D., C.W. Engelbracht, D. Fadda, J.A. Stansberry, and S. Wachter 2007. Absolute calibration and characterization of the Multiband Imaging Photometer for Spitzer. II. 70 micron imaging. *Publ. Astron. Soc. Pacific* (submitted).

Grundy, W.M., K.S. Noll, and D.C. Stephens 2005. Diverse albedos of small transneptunian objects. *Icarus* **176,** 184-191.

Holman, M.J. 1997. A possible long-lived belt of objects between Uranus and Neptune. *Nature* **387,** 785-788.

Horner, J., N.W. Evans, and M.E. Bailey 2004. Simulations of the population of Centaurs - I. The bulk statistics. *Mon. Not. Royal Astron. Soc.* **354,** 798-810.

Hubbard, W.B. 1984. *Planetary interiors.* Van Nostrand Reinhold Co., New York NY.

Jenniskens, P, D.F. Blake, and K. Kouchi 1998. Amorphous water ice: A solar system material. In *Solar System Ices*, Eds. B. Schmitt, C. de Bergh, and K. Kouchi, Kluwer Academic Publishers, Boston MA.

Jewitt, D., J. Luu, and J. Chen 1996. The Mauna Kea-Cerro-Tololo (MKCT) Kuiper belt and Centaur survey. *Astron. J.* **112,** 1225-1238.

Jewitt, D., L. Chizmadia, R. Grimm, and D. Prialnik 2007. Water in Small Bodies of the Solar System. In *Protostars and Planets V*, Eds. B. Reipurth, D. Jewitt, and K. Keil, Univ. of Arizona Press, Tucson AZ.

Johnson, T.V., and J.I. Lunine 2005. Saturn's moon Phoebe as a captured body from the outer Solar System. *Nature* **435,** 69-71.

Kern, S.D., and J.L. Elliot 2006. The frequency of binary Kuiper belt objects. *Astrophys. J.* **643,** L57-L60.

Kowal, C.T., W. Liller, and B.G. Marsden 1979. The discovery and orbit of 2060 Chiron. *IAU Symp. 81: Dynamics of the Solar System* **81,** 245-250.

Krist, J.E., and R.N. Hook 2004. *The Tiny Tim user's guide.* Version 6.3 is available from `http://www.stsci.edu/software/tinytim/`.

Lacerda, P., and D.C. Jewitt 2007. Densities of Solar System objects from their rotational lightcurves. Astron. J. **133,** 1393-1408.

Lebofsky, L.A., and J.R. Spencer 1989. Radiometry and thermal modeling of asteroids. In *Asteroids II*, Eds. R.P. Binzel, T. Gehrels, and M.S. Matthews, Univ. of Arizona Press, Tucson AZ.

Leliwa-Kopystyński, J., and K.J. Kossacki 2000. Evolution of porosity in small icy bodies. *Planet. Space Sci.* **48,** 727-745.

Levison, H.F., and M.J. Duncan 1997. From the Kuiper belt to Jupiter-family comets: The spa-

tial distribution of ecliptic comets. *Icarus* **127,** 13-32.

Lunine, J.I., and W.C. Tittemore 1993.  Origins of outer-planet satellites.  In *Protostars and Planets III*, Eds. E.H. Levy, and J.I. Lunine, Univ. of Arizona Press, Tucson AZ.

Lupo, M.J., and J.S. Lewis 1979.  Mass-radius relationships in icy satellites. *Icarus* **40,** 157-170.

Marchis, F., D. Hestroffer, P. Descamps, J. Berthier, A.H. Bouchez, R.D. Campbell, J.C.Y. Chin, M.A. van Dam, S.K. Hartman, E.M. Johansson, R.E. Lafon, D. Le Mignant, I. de Pater, P.J. Stomski, D.M. Summers, F. Vachier, P.L. Wizinovich, and M.H. Wong 2006.  A low density of 0.8 g cm$^{-3}$ for the Trojan binary asteroid 617 Patroclus. *Nature* **439,** 565-567.

McKinnon, W.B., J.I. Lunine, and D. Banfield 1995.  Origin and evolution of Triton.  In *Neptune and Triton*, Ed. D.P. Cruikshank, Univ. of Arizona Press, Tucson AZ.

McKinnon, W.B., W.B. Durham, and L.A. Stern 2005.  Cold compaction of porous ice, and the densities of Kuiper belt objects.  Paper presented at *Asteroids, Comets, and Meteors* 2005 August 7-12, Rio de Janeiro, Brazil.

Nelder, J., and R. Mead 1965.  A simplex method for function minimization. *Computer J.* **7,** 308-313.

Noll, K.S., D.C. Stephens, W.M. Grundy, and I. Griffin 2004a.  The orbit, mass, and albedo of (66652) 1999 RZ$_{253}$. *Icarus* **172,** 402-407.

Noll, K.S., D.C. Stephens, W.M. Grundy, D.J. Osip, and I. Griffin 2004b.  The orbit and albedo of trans-neptunian binary (58534) 1997 CQ$_{29}$. *Astron. J.* **128,** 2547-2552.

Noll, K.S., H.F. Levison, W.M. Grundy, and D.C. Stephens 2006.  Discovery of a binary Centaur. *Icarus* **184,** 611-618.

Noll, K.S., W.M. Grundy, E. Chiang, and J.L. Margot 2007.  Binaries in the Kuiper belt.  In *Kuiper Belt*.  Eds. A. Barucci, H. Boehnhardt, D. Cruikshank, and A. Morbidelli. Univ. of Arizona Press, Tucson AZ (in press).

Peixinho, N., A. Doressoundiram, A. Delsanti, H. Boehnhardt, M.A. Barucci, and I. Belskaya 2003.  Reopening the TNOs color controversy: Centaurs bimodality and TNOs unimodality. *Astron. & Astrophys.* **410,** L29-L32.

Petit, J.M., and O. Mousis 2004.  KBO binaries: How numerous are they? *Icarus* **168,** 409-419.

Press, W.H., S.A. Teukolsky, W.T. Vetterling, and B.P. Flannery 1992. *Numerical Recipes in C*. Cambridge Univ. Press, New York NY.

Rabinowitz, D.L., K. Barkume, M.E. Brown, H. Roe, M. Schwartz, S. Tourtellotte, and C. Trujillo 2006.  Photometric observations constraining the size, shape, and albedo of 2003 EL$_{61}$, a rapidly rotating, Pluto-sized object in the Kuiper belt. *Astrophys. J.* **639,** 1238-1251.

Rasio, F.A., C.A. Tout, S.H. Lubow, and M. Livio 1996, Tidal decay of close planetary orbits. *Astrophys. J.* **470,** 1187-1191.

Rieke, G.H., and 42 co-authors 2004.  The Multiband Imaging Photometer for Spitzer (MIPS). *Astrophys. J.* **154,** 25-29.

Rousselot, P., J.M. Petit, F. Poulet, and A. Sergeev 2005.  Photometric study of Centaur (60558)

2000 EC$_{98}$ and trans-neptunian object (55637) 2002 UX$_{25}$ at different phase angles. *Icarus* **176,** 478-491.

Schubert, G., T. Spohn, and R.T. Reynolds 1986. Thermal histories, compositions, and internal structures of the moons of the solar system. In *Satellites*, Univ. of Arizona Press, Tucson AZ.

Sheppard, S.S., and D.C. Jewitt 2002. Time-resolved photometry of Kuiper belt objects: Rotations, shapes, and phase functions. *Astron. J.* **124,** 1757-1775.

Spencer, J.R., J.A. Stansberry, W.M. Grundy, and K.S. Noll 2006. A low density for binary Kuiper belt object (26308) 1998 SM$_{165}$. *Bull. Amer. Astron. Soc.* **38,** 546 (abstract).

Sirianni, M., M.J. Jee, N. Benítez, J.P. Blakeslee, A.R. Martel, G. Meurer, M. Clampin, G. De Marchi, H.C. Ford, R. Gilliland, G.F. Hartig, G.D. Illingworth, J. Mack, and W.J. McCann 2005. The photometric performance and calibration of the HST Advanced Camera for Surveys. *Publ. Astron. Soc. Pacific* **117,** 1049-1112.

Stansberry, J.A., W.M. Grundy, J.L. Margot, D.P. Cruikshank, J.P. Emery, G.H. Rieke, and D.T. Trilling 2006. The albedo, size, and density of binary Kuiper belt object (47171) 1999 TC$_{36}$. *Astrophys. J.* **643,** 556-566.

Stansberry, J., W. Grundy, M. Brown, D. Cruikshank, J. Spencer, D. Trilling, and J.L. Margot 2007a. Physical properties of Kuiper belt objects and Centaurs: Constraints from Spitzer Space Telescope. In *Kuiper Belt*. Eds. A. Barucci, H. Boehnhardt, D. Cruikshank, and A. Morbidelli. Univ. of Arizona Press, Tucson AZ (in press).

Stansberry J.A., K.D. Gordon, B. Bhattacharya, C.W. Engelbracht, G.H. Rieke, M. Blaylock, F.R. Marleau, D. Fadda, D.T. Frayer, A. Noriega-Crespo, S. Wachter, D. Henderson, D.M. Kelly, and G. Neugebauer 2007b. Absolute calibration and characterization of the Multiband Imaging Photometer for Spitzer III. An asteroid-based calibration at 160 microns. *Publ. Astron. Soc. Pacific* (submitted).

Stephens, D.C., and K.S. Noll 2006. Detection of six trans-neptunian binaries with NICMOS: A high fraction of binaries in the cold classical disk. *Astron. J.* **131,** 1142-1148.

Tegler, S.C., W. Romanishin, and G.J. Consolmagno, S.J. 2003. Color patterns in the Kuiper belt: A possible primordial origin. *Astrophys. J.* **599,** L49-L52.

Trilling, D.E. 2000. Tidal constraints on the masses of extrasolar planets. *Astrophys. J.* **537,** L61-L64.

Trilling, D.E. 2007. The evolution of Kuiper belt object and Centaur binaries. *Astrophys. J.* (submitted).

Trilling, D.E., and G.M. Bernstein 2006. Lightcurves of 20-100 kilometer Kuiper belt objects using the Hubble Space Telescope. *Astron. J.* **131,** 1149-1162.

Veillet, C., J.W. Parker, I. Griffin, B. Marsden, A. Doressoundiram, M. Buie, D.J. Tholen, M. Connelley, and M.J. Holman 2002. The binary Kuiper-belt object 1998 WW$_{31}$. *Nature* **416,** 711-713.

Virtanen, J., K. Muinonen, and E. Bowell 2001. Statistical ranging of asteroid orbits. *Icarus* **154,**

412-431.

Virtanen, J., G. Tancredi, K. Muinonen, and E. Bowell 2003. Orbit computation for trans-neptunian objects. *Icarus* **161,** 419-430.

Yoder, C.F. 1982. Tidal rigidity of Phobos. *Icarus* **49,** 327-346.

**Figure 5.**